\documentclass[iop]{emulateapj}
\usepackage{graphicx}
\usepackage{natbib}
\bibpunct{(}{)}{;}{a}{}{,}
\submitted{Received 2010 September 27; accepted 2011 February 3}
\begin{document}

\title{Detection of X-rays from the symbiotic star V1329 Cyg}

\author{Matthias Stute\altaffilmark{1,2}}
\altaffiltext{1}{Institute for Astronomy and Astrophysics, Section Computational
Physics, Eberhard Karls Universit\"at T\"ubingen, Auf der Morgenstelle 10, 
72076 T\"ubingen, Germany}
\altaffiltext{2}{Dipartimento di Fisica Generale "A. Avogadro", Universit\`a 
degli Studi di Torino, Via Pietro Giuria 1, 10125 Torino, Italy}
\email{matthias.stute@tat.physik.uni-tuebingen.de}
\author{Gerardo J. M. Luna\altaffilmark{3}}
\altaffiltext{3}{Harvard-Smithsonian Center for Astrophysics, 60 Garden St.
MS 15, Cambridge, MA, 02138, USA}
\author{Jennifer L. Sokoloski\altaffilmark{4}}
\altaffiltext{4}{Columbia Astrophysics Laboratory, 550 W. 220th Street, 1027
Pupin Hall, Columbia University, New York, NY 10027, USA}

\begin{abstract}
We report the detection of X-ray emission from the symbiotic star V1329\,Cyg 
with XMM-Newton. The spectrum from the EPIC pn, MOS1 and MOS2 instruments 
consists of a two-temperature plasma with $k\,T_1 = 0.11^{+0.02}_{-0.02}$ keV and
$k\,T_2 = 0.93^{+0.12}_{-0.14}$ keV. Unlike the vast majority of symbiotic stars 
detected in X-rays, the soft component of the spectrum seems to be absorbed 
only by interstellar material. The shock velocities corresponding to the 
observed temperatures are about 300 km s$^{-1}$ and about 900 km s$^{-1}$. 
We did not find either periodic or aperiodic X-ray variability, with
upper limits on the amplitudes of such variations being 46\% and 16\%
(rms), respectively.  We also did not find any ultraviolet variability
with an rms amplitude of more than approximately 1\%. The derived velocities 
and the unabsorbed nature of the soft component of the X-ray spectrum suggest 
that some portion of the high energy emission could originate in shocks within a
jet and beyond the symbiotic nebula. The lower velocity is consistent with the 
expansion velocity of the extended structure present in HST observations. The 
higher velocity could be associated with an internal shock at the base of the 
jet or with shocks in the accretion region.
\end{abstract}


\maketitle

\section{Introduction}

V1329\,Cygni ($=$HBV 475) is one of a small number of symbiotic novae 
\citep{MuN94} that have outbursts of 3 to 7 mag and return to their previous 
brightness only slowly over decades. The canonical model \citep[e.g.][]{BaV90} 
for symbiotic novae involves a hot white dwarf star that has a surface 
thermonuclear flash, with the fuel supplied by a red giant companion. The only 
recorded major outburst of V1329 Cyg began in about 1965 and reached B 
magnitude $<11.5$ mag in October 1966. Some time elapsed between discovery as an
eruptive object \citep{Koh69,KoB70} and recognition that it is a binary, with 
the object first being called a proto-planetary nebula \citep{CGK70}. The 
photographic brightness dropped to $m_V<16$ mag over the past century, with 
several abrupt drops of as much as another 2.5 mag between 1925 and 1962. These 
repeated with a period of 950--959 days and were explained as the eclipses of 
the hot component by the red giant \citep{SCS74, GHC79}. \citet{ScS97} improved 
the determination of the orbital period to 956.5 days and found from polarimetry
an inclination of $86\pm2$ degrees. From optical and UV emission lines, orbital 
parameters of the hot component have been determined yielding minimum masses of 
about 0.71 and 2 M$_{\odot}$ for the white dwarf and red giant, respectively 
\citep{ScS97}. These authors derived an $E_{B-V}$ value of 0.6, corresponding to 
$n_H=2.3\times10^{21}$ cm$^2$ \citep[using the conversion factor of][]{GrL89}.

About 200 symbiotic stars are known \citep[e.g.][]{BMM00}, but jets have been 
detected at different wavelengths only in 10 of them \citep{BSK04}. 
V1329\,Cygni is a member of this list, since \citet{BBE03} found two peaks and 
extended emission in HST WFPC2 snapshot images taken in October 1999 with F502N 
and F656N filters. These were separated by about 950 AU, assuming a distance of 
3.4 kpc as given by \citet{MuN94}. After comparison with HST images of 
\citet{ScS97} taken in July 1991, they derived an expansion velocity of 
$260\pm50$ km s$^{-1}$, suggesting that this mass ejection was not associated 
with the nova outburst in 1965, but with an event in 1982. As an additional 
riddle, the position angle of the orbital plane of $11\pm2^\circ$ determined by 
\citet{ScS97} furthermore suggests that the mass ejection occurred {\em along} 
the orbital plane, instead of perpendicular to it as expected for a jet. Using 
the ephemeris of \citet{ScS97}
\begin{equation}
\textrm{JD$_{\rm min}$} = 2 444 890.0 + 956.5 \times E\, ,
\end{equation}
the system was at phase 0.733 during the observations of \citet{ScS97} and at 
phase 0.880 during that of \citet{BBE03}. Whether the different appearance of 
the extended [OIII] emission is due to this slightly different orbital phase 
and thus a changed ionisation structure in the stellar environment, or due to 
real expansion of the gas inside a wind or jet, is unknown. We also
have to note that \citet{ScS97} used the FOC camera with filter F501N, while 
\citet{BBE03} used the WFPC2 camera with filter F502N, thus different filters 
with different properties.

X-ray observations provide a direct probe of the two most important components 
of jet-driving systems: the bow and internal shocks of the jet emitting soft 
X-rays and the central parts of the jet engine, where gas is being accreted to 
power the jet, leading to hard and/or super-soft X-ray emission. Currently R\,Aqr 
\citep{KPL01,KAK07} and CH\,Cyg \citep{GaS04, KCR07,KGC10} are the only two jets
from symbiotic stars that have been resolved in X-rays. All objects with jets, 
detected in other wavelength, when observed in X-rays, show soft components with
$k\,T<2$ keV (additionally to sometimes present {\em super-soft} emission): R 
Aqr \citep{KPL01,KAK07}, CH Cyg \citep{GaS04, KCR07,KGC10}, MWC 560 
\citep{StS09}, RS Oph \citep{LMS09}, AG Dra \citep{GVI08}, Z And \citep{SKE06}. 
The three objects CH Cyg, R Aqr, MWC 560 also emit hard components 
\citep{MIK07,NDK07,StS09}. Z And showed hard emission in one of three
observations only \citep{SKE06}.

The paper is organized as follows: in Section \ref{sec_obs}, we show details of 
the observations and the analysis of the data. After that we describe the 
results in Section \ref{sec_res}. We end with a discussion and conclusions in 
Sections \ref{sec_dis} and \ref{sec_concl}. 

\section{Observation and analysis} \label{sec_obs}

We observed the field of V1329\,Cyg with {\em XMM-Newton} in 2009 
(Table \ref{Tbl_obs}) using the EPIC instrument operating in full window mode 
and with the medium thickness filter. Simultaneously, we used the Optical 
Monitor OM. All the data reduction was performed using the Science Analysis 
Software (\textsc{SAS}) software 
package\footnote{\url{http://xmm.vilspa.esa.es/}}
version 8.0. We removed event at periods with high background levels from the 
pipeline products selecting events with pattern 0--4 (only single and double 
events) for the pn and pattern 0--12 for the MOS, respectively, and applying 
the filter {FLAG=0}. The resulting exposure time after these steps is 37.6 ks. 

\begin{deluxetable}{lllll}[!h]
\tablecaption{Observations on May 02/03, 2009 (ObsID: 0604920301) 
\label{Tbl_obs}}
\tablewidth{\columnwidth}
\tablecolumns{5}
\tablehead{\colhead{Instrument} & \colhead{Filter} & \colhead{Duration} & 
\colhead{UT Start} & \colhead{UT Stop}}
\startdata
pn   & Medium & 57434                 & 11:00:56 & 02:58:10 \\
MOS1 & Medium & 59021                 & 10:38:33 & 03:02:14 \\
MOS2 & Medium & 59027                 & 10:38:33 & 03:02:20 \\
\tableline
OM   & U      & 10$\times$ $\sim$1500 & 10:47:01 & 16:15:16 \\
     & UVW1   & 10$\times$ $\sim$1500 & 16:50:37 & 21:48:52 \\
     & UVM2   & 10$\times$ $\sim$1500 & 21:54:13 & 03:04:49 
\enddata
\end{deluxetable}

The source spectra and light curves were accumulated from a circular region of 
12'' radius (240 pixels, Fig. \ref{fig_images}) centered on V1329\,Cyg using 
the coordinates from \textsc{SIMBAD}. The background spectra and light curves 
were extracted from a source-free region on the same chips taken within an 
circle of 30'' radius (600 pixels). Spectral redistribution matrices and 
ancillary response files were generated using the \textsc{SAS} scripts 
\texttt{rmfgen} and \texttt{arfgen}, and spectra grouped with a minimum of 25 
counts per energy bin were fed into the spectral fitting package 
\textsc{XSPEC}\footnote{\url{http://heasarc.gsfc.nasa.gov/docs/xanadu/xspec/}} 
v12.5.1. For timing analysis, photon arrival times were converted to the solar 
system barycenter using the SAS task \texttt{barycen}. 

\section{Results} \label{sec_res}

\subsection{Images}

\begin{figure*}[!htb]
  \centering
  \includegraphics[scale=0.24]{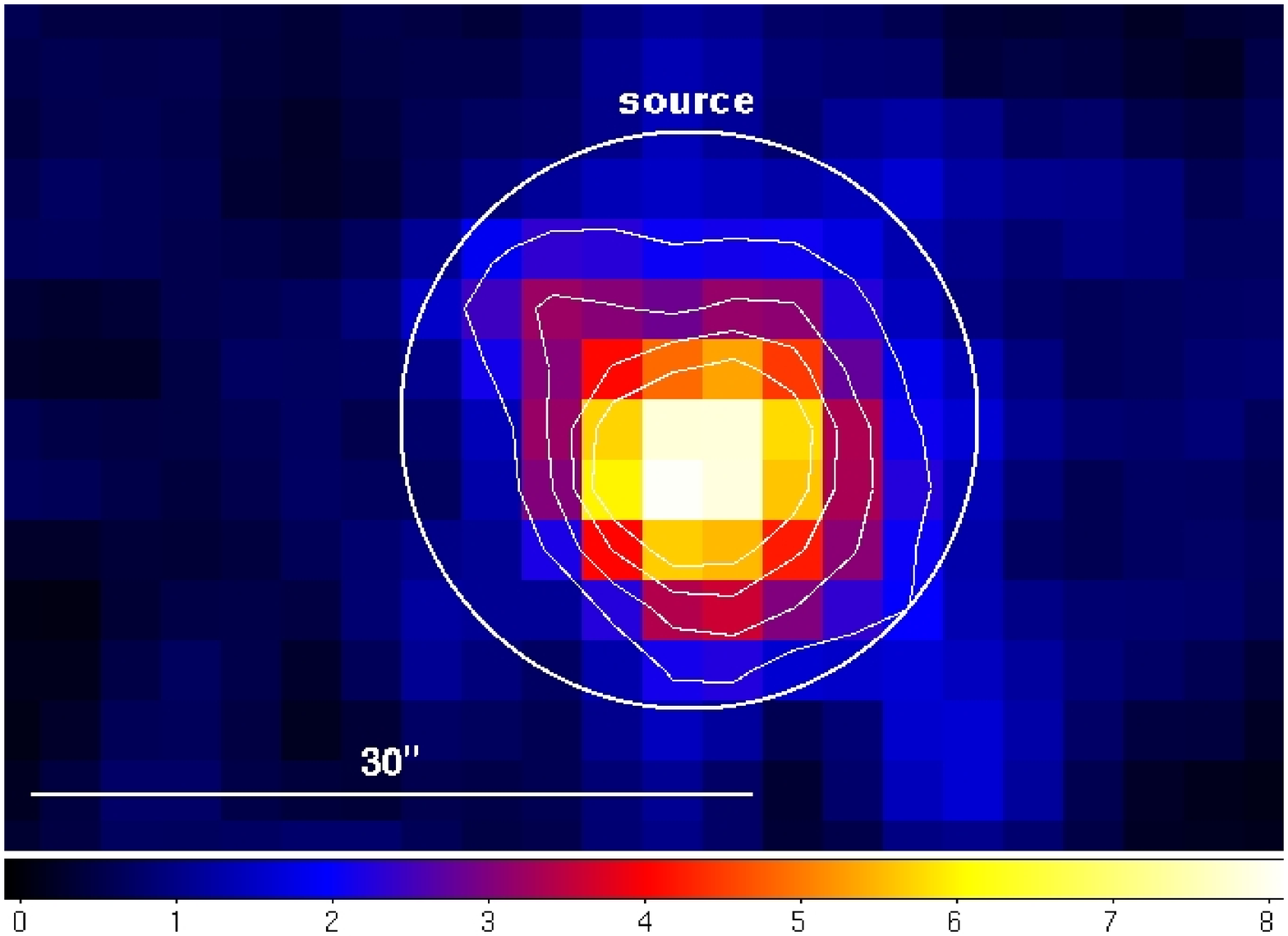}
  \includegraphics[scale=0.24]{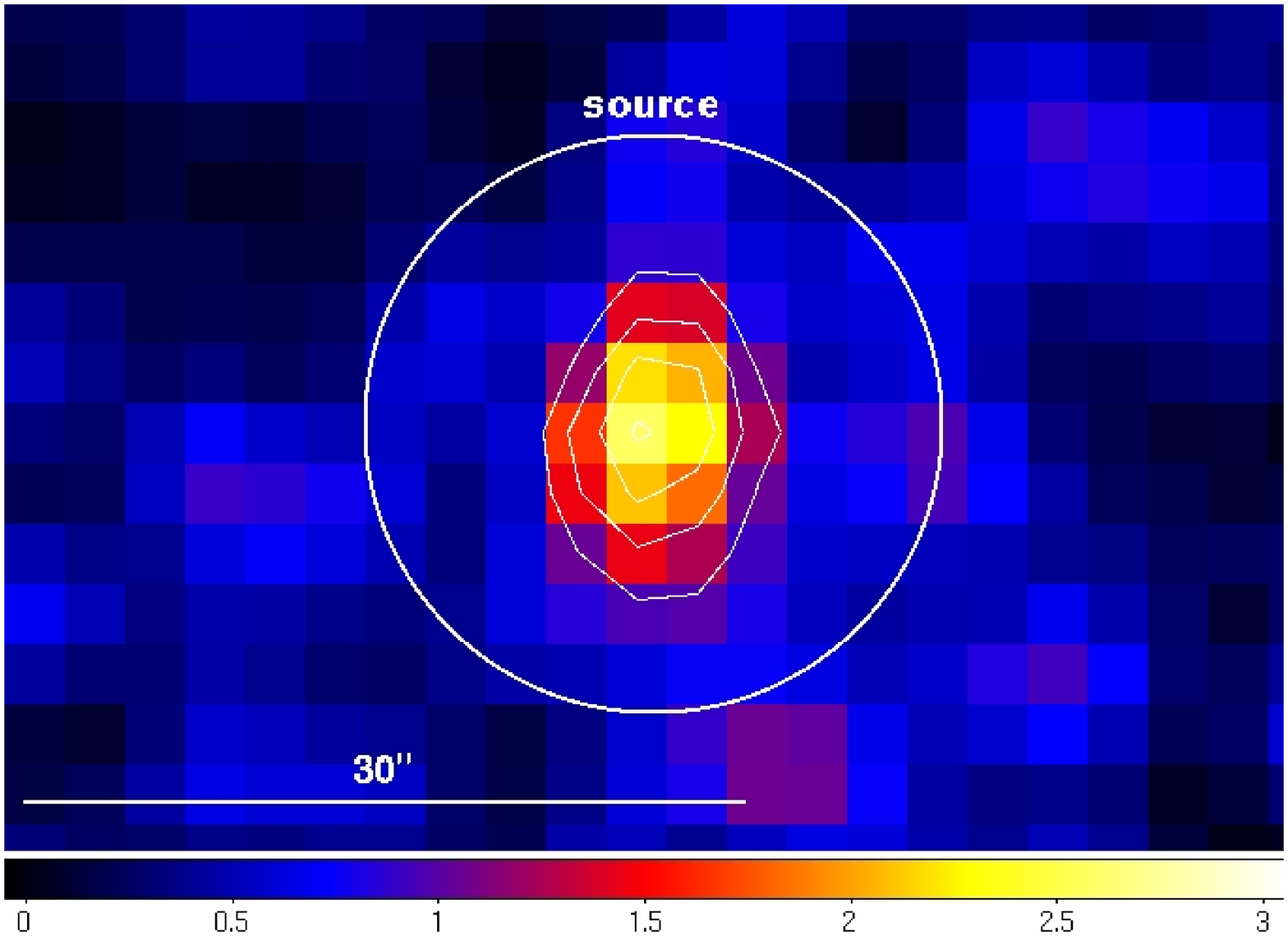}
  \includegraphics[scale=0.24]{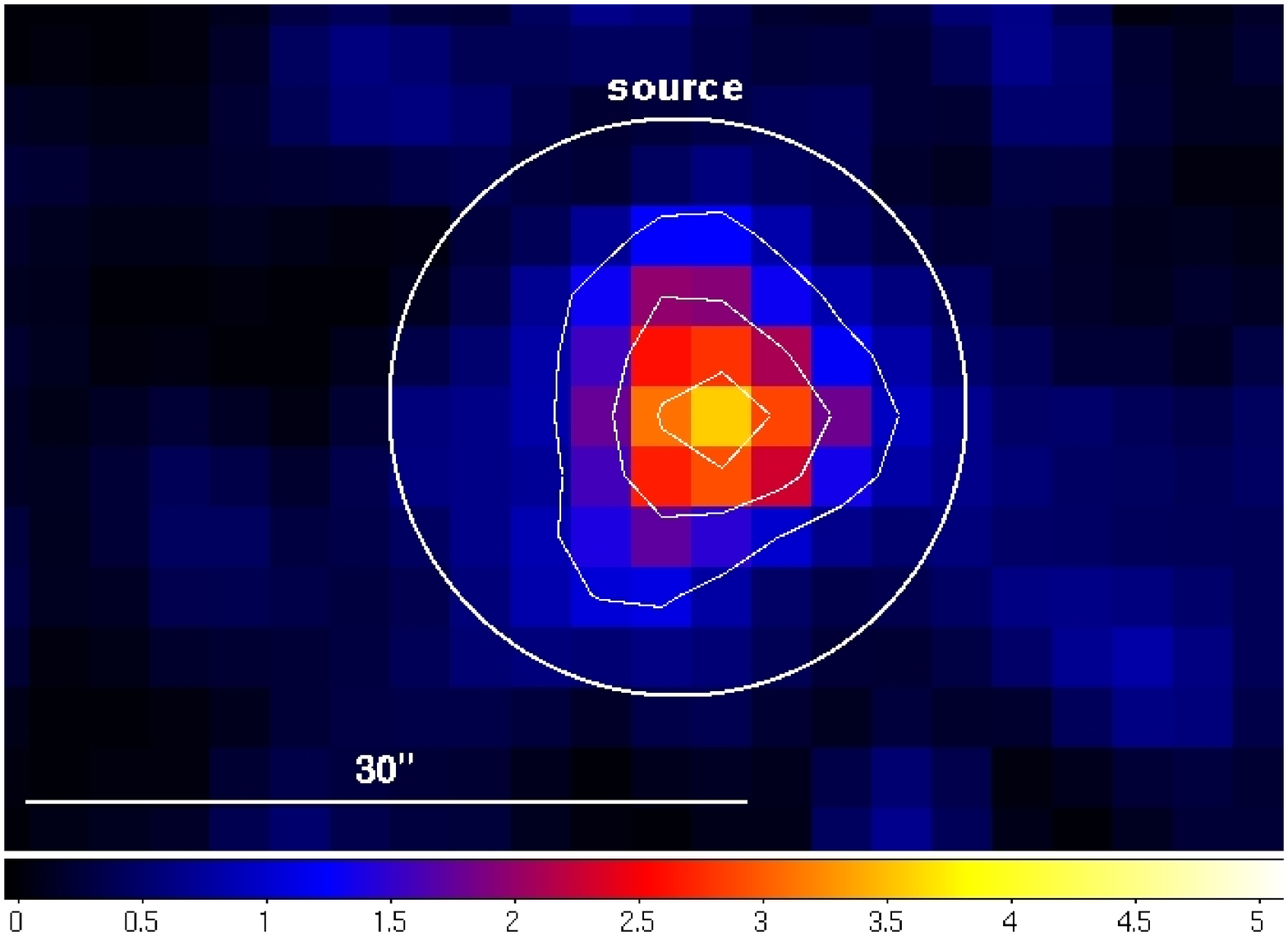}
  \caption{Left: {\em XMM-Newton} EPIC pn image (detail) of the region around 
    V1329\,Cyg in the 0.3--2 keV range. Also shown is the extraction region 
    for the source events. Contour levels are 2, 3, 4 and 5 counts per pixel; 
    middle: {\em XMM-Newton} EPIC MOS1 image of the same region. Contour levels 
    are 1.5, 2 and 2.5 counts per pixel; right: {\em XMM-Newton} EPIC MOS2 
    image of the same region. Contours are 1, 2, 3 and 4 counts per pixel.}
  \label{fig_images}
\end{figure*}

We detected V1329 Cyg with each of the instruments aboard XMM-Newton.  The 
total numbers of counts in the EPIC pn, MOS1, and MOS2 cameras were 329, 84, 
and 107, respectively (Fig. \ref{fig_images}). For comparison, the total numbers
counts expected due to background radiation alone were 83, 36, and 20 for the 
three X-ray instruments. The average magnitudes in the optical filters U, UVW1 
and UVM2 are 12.32, 12.34 and 13.35, respectively. The X-ray images of 
V1329\,Cyg show a point source morphology. In order to assess the possibility of
extended emission, we constructed radial profiles from the images using the 
\textsc{SAS} task \texttt{eradial} (a routine to extract a radial profile of a 
source in an image field and to fit a point spread function [PSF] to it) keeping
the centroid fixed at the source position of V1329 Cyg and choosing a step size 
of two pixels. With this method, we did not find any evidence for extended X-ray
emission.

\subsection{Spectra}

\begin{figure}[!htb]
  \centering
  \includegraphics[height=\columnwidth,angle=-90]{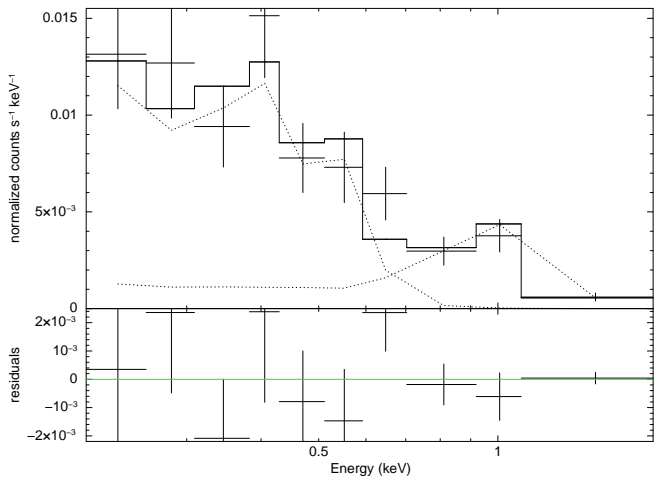}
  \caption{Observed spectrum of V1329\,Cyg together with a model consisting of 
    two absorbed thermal components (\texttt{vapec}; dotted lines) with 
    variable abundances.}
  \label{fig_spectralfit1}
\end{figure}

The spectrum of V1329\,Cyg is relatively soft with photons having energies of 
$<2$ keV (Fig. \ref{fig_spectralfit1}). It is well described ($\chi^2 = 1.10$, 
6 d.o.f.) by a two-temperature plasma model with N overabundant by a factor of 
$\approx 4$ in the softer component compared to solar values \citep{AnG89} in 
order to explain the high flux around 0.4 keV (\texttt{vapec} in 
\textsc{XSPEC}). For the soft component, the fit is consistent with no 
{\em internal} absorption, with $N_H$ in the range $(0-5)\times10^{21}$ 
cm$^{-2}$. For the hard component, we find $N_H = 9^{+4.3}_{-0.32}\times10^{21}$ 
cm$^{-2}$. These values include internal absorption as well as interstellar 
absorption which in the direction of the source is about $2.3\times10^{21}$ 
cm$^{-2}$. The fit is slightly worse ($\chi^2 = 1.27$, 4 d.o.f.) including 
(non-vanishing) absorption. The temperature of the cooler component is 
$k\,T_1 = 0.11^{+0.02}_{-0.02}$ keV, while the temperature of the hotter component 
is $k\,T_2 = 0.93^{+0.12}_{-0.14}$ keV. The flux of the hotter component is three 
times lower than that of the soft component. The total model flux between 0.3 
and 2 keV is $1.21\times10^{-14}$ erg s$^{-1}$\,cm$^{-2}$. With the assumed 
distance of 3.4 kpc, this flux corresponds to a luminosity of 
$1.7\times10^{31}$ erg s$^{-1}$.

\begin{figure}[!htb]
  \centering
  \includegraphics[width=\columnwidth]{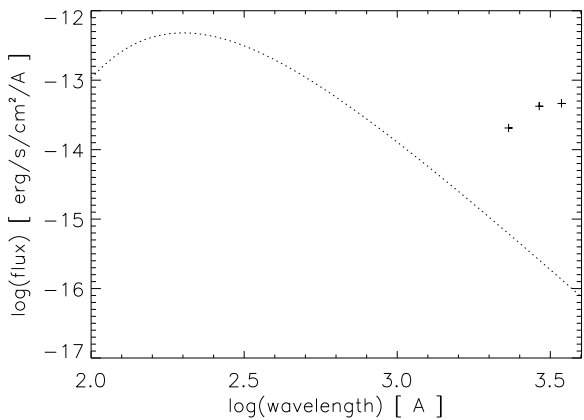}
  \caption{Spectral energy distribution in the optical filters U, UVW1 
    and UVM2. Also plotted is blackbody emission representing a white dwarf 
    with a temperature of 145000 K and radius of $8\times10^8$ cm. The slope 
    and flux level of the spectral energy distribution indicate that the 
    optical flux is dominated by nebular emission and the red giant.}
  \label{fig_SED}
\end{figure}

The slope of the ultraviolet spectral energy distribution indicates that the 
optical flux is dominated by nebular emission and the red giant (Fig. 
\ref{fig_SED}). The flux level of blackbody emission corresponding to a white 
dwarf with a radius of $8\times10^8$ cm at a distance of 3.4 kpc with an 
effective temperature of 145000 K \citep{MNS91} is lower than the measured 
fluxes by 1--2 orders of magnitude.

\subsection{Light curves}

\begin{figure}[!htb]
  \centering
  \includegraphics[width=\columnwidth]{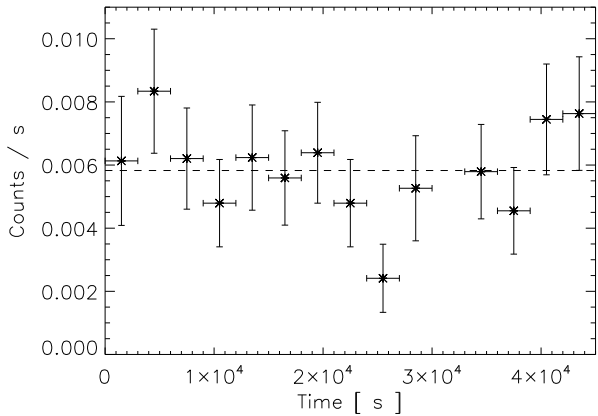}
  \caption{EPIC pn light curve binned at 3000 s in the energy range 0.3--2 keV. 
  The dashed line shows the average count rate. Overall the light curve is 
  consistent with X-ray flux being constant -- the ratio of measured rms to that
  expected from Poisson statistics,  $s / s_{exp}$, is 1.03.}
  \label{fig_lightcurve_PN}
\end{figure}

\begin{figure}[!htb]
  \centering
  \includegraphics[width=\columnwidth]{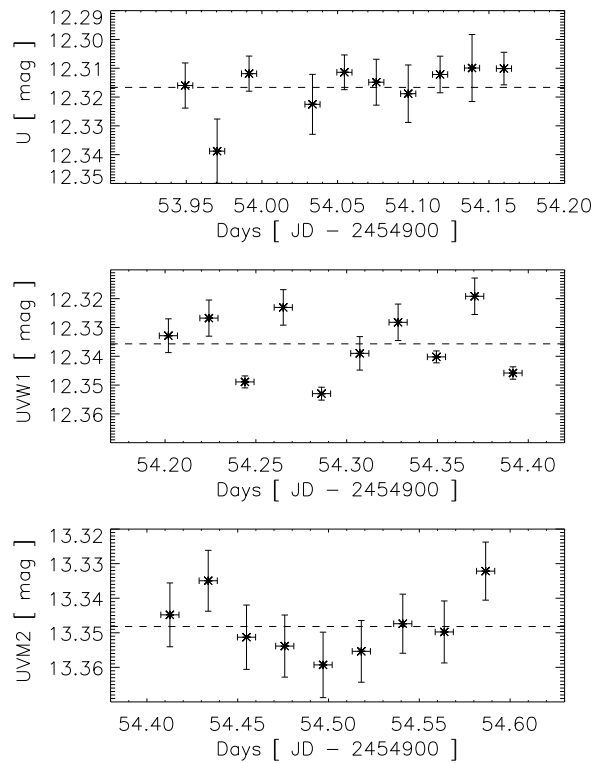}
  \caption{Light curves of the OM exposures for filters U (top, 3440 \AA), UVW1 
    (middle, 2910 \AA) and UVM2 (bottom, 2310 \AA). The dotted lines show the 
    average magnitude. The light curves for filters U and UVM2 are consistent 
    with a constant flux ($s / s_{exp} = 0.89$ and $0.94$), while 
    $s / s_{exp} = 3.77$ for the UVW1 filter indicates variability.}
  \label{fig_lightcurve_OM}
\end{figure}

We examined time series binned at 3000s in the energy range 0.3--2 keV (Fig. 
\ref{fig_lightcurve_PN}) and the OM photometry with exposure times of about 
1500s (Fig. \ref{fig_lightcurve_OM}). We chose the bin size for the X-ray light 
curve to produce at least roughly 25 counts per bin while still providing time
resolution of less than an hour. The X-ray and two of the optical light 
curves are consistent with a constant flux. The measured rms variations, those 
expected from Poisson statistics, and the ratios of these two quantities are 
listed in Table \ref{Tbl_rms}. $s$ and $s_{exp}$ are given in percentage of the 
mean value. Only in filter UVW1, the ratio $s / s_{exp} > 1$ indicates 
variability. 

\begin{deluxetable*}{lllll}[!htb]
\tablecaption{Measured and expected variations and their ratio
\label{Tbl_rms}}
\tablewidth{\columnwidth}
\tablecolumns{4}
\tablehead{\colhead{Instrument} & \colhead{measured variation $s$} & 
\colhead{expected variation $s_{exp}$} & \colhead{ratio $s / s_{exp}$}}
\startdata
X-ray        & 39.62 \% & 38.56 \% & 1.03 \\
optical U    & 0.054 \% & 0.061 \% & 0.89 \\
optical UVW1 & 0.095 \% & 0.025 \% & 3.77 \\
optical UVM2 & 0.063 \% & 0.067 \% & 0.94 
\enddata
\end{deluxetable*}

We also searched for modulated emission in the light curves. Using the arrival 
times of the source events, we calculated the $Z^2_1$ (Rayleigh) statistic 
\citep{BBB83} in the frequency range $f_{\rm min} = 1/T_{\rm exp}$ to 
$f_{\rm max} = 1/2\,t_{\rm frame}$ with $\Delta\,f = 1/T_{\rm exp}$, where
$T_{\rm exp}$ is the effective exposure time of 37.6 ks and  $t_{\rm frame}$ is 
the readout time (73 ms in the case of XMM). The value of $Z^2_1$ needed to 
detect pulsation with a probability $P$ is given by
\begin{equation}
Z^2_1 > 2\times \ln (\frac{T_{\rm exp} \times \Delta f}{P})\,.
\end{equation}
Thus for a 3-$\sigma$ ($P = 2.699 \times 10^{-3}$) detection, $Z^2_1 > 36.7$ is
required. The highest value of $Z^2_1$ detected in our observation is 27.5, 
which corresponds only to 1.26-$\sigma$. With $N_{S}$=329 and $N_{B}$=83 source 
and background photons detected with the EPIC pn camera respectively, the pulse 
fraction corresponding to the highest $Z_1^2$ value detected is 
\begin{eqnarray}
p &=& p_{obs}(N_{S}+N_{B})N^{-1}_{S} \nonumber \\
&\approx& [2Z_{1}^{2} (N_{S}+N_{B})]^{1/2}N^{-1}_{S} 
\pm [2(N_{S}+N_{B})]^{1/2}N^{-1}_{S} \nonumber \\
&=& 0.46\pm0.009
\end{eqnarray}
for a nearly sinusoidal signal, where $p_{obs}$ is the pulsed fraction 
uncorrected for the background contribution. Therefore, our observation is only 
sensitive to pulse fractions $p \gtrsim 46$ \%.

\section{Discussion} \label{sec_dis}

\begin{figure*}[!htb]
  \centering
  \includegraphics[height=0.45\textwidth]{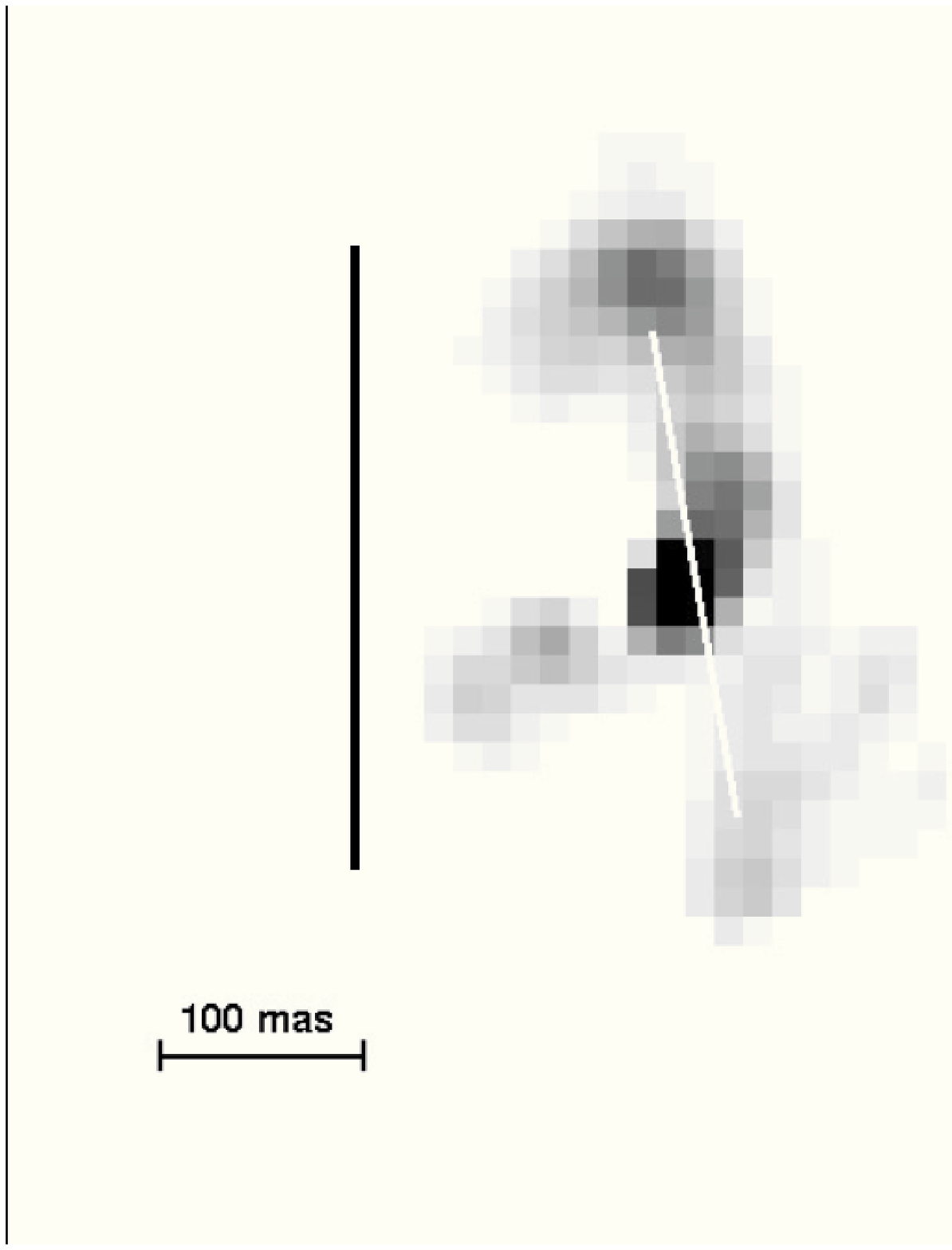}
  \includegraphics[height=0.45\textwidth]{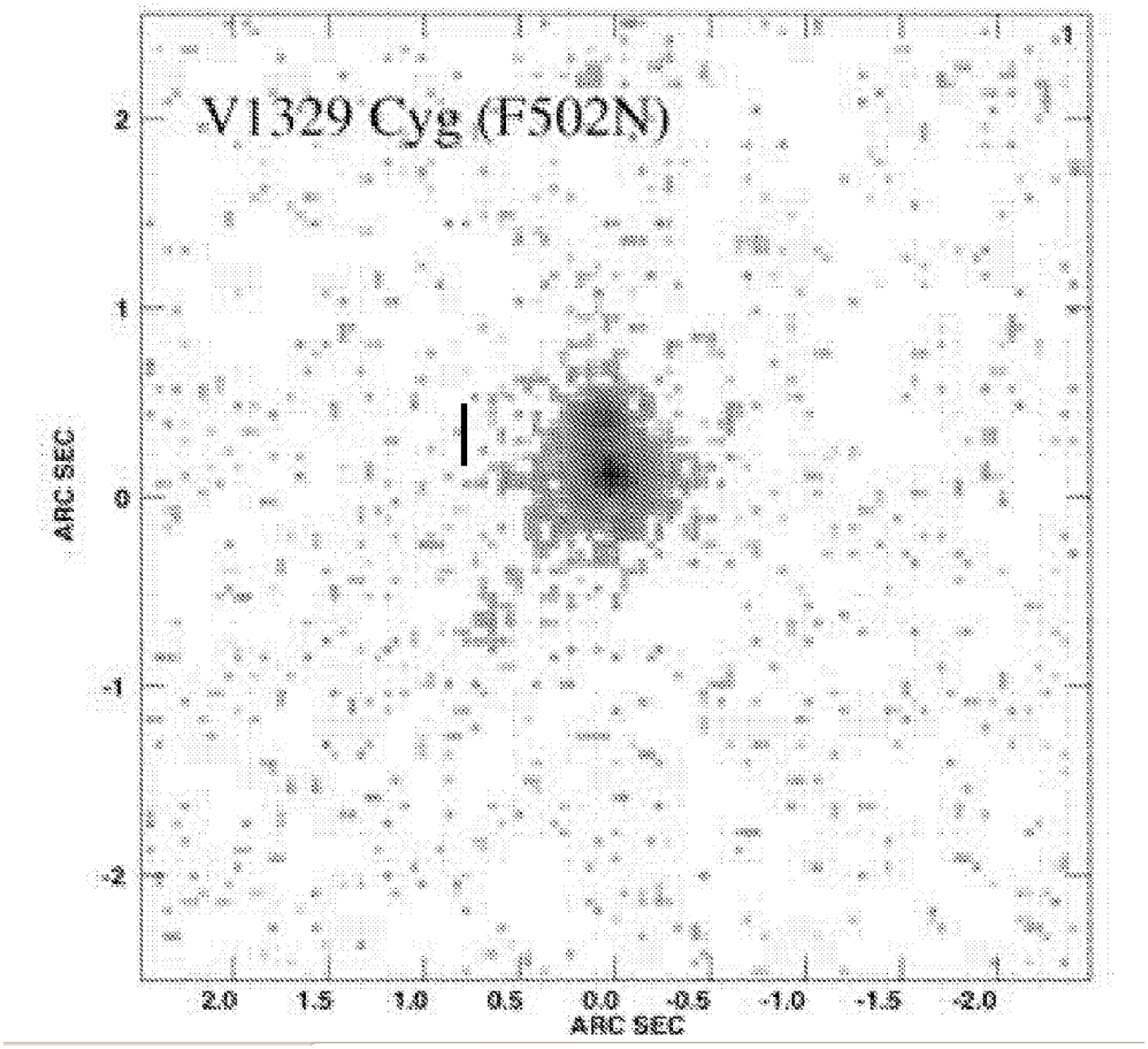}
  \caption{Left: {\em HST} image in the filter F501N of the 
    region around V1329\,Cyg from \citet{ScS97}. The orientation of the orbital 
    plane is indicated by the white line; right: {\em HST} image in the filter 
    F502N of the same region from \citet{BBE03}. For comparison, a scale of 300 
    mas has been added in both images.}
  \label{fig_images_HST}
\end{figure*}

The X-ray images of V1329\,Cyg show a point source morphology; no hints of 
extended emission are present. In Figure \ref{fig_images_HST}, we show 
{\em HST} images from \citet{ScS97} and \citet{BBE03} where extended emission 
in [OIII] is evident. These authors concluded that the visible optical emission 
is an expanding jet, however, resulting in the conundrum that the mass ejection 
occurred {\em along} the orbital plane. 

The two-temperature X-ray emission can be explained in the light of simulations 
of jets in symbiotic stars \citep{Stu06,StS07} where soft X-ray emission arises 
from internal shocks and the bow shock. The velocity of the shock, 
$v_{\rm shock}$, can be derived using the measured temperature of the 
component and the following relation (assuming strong shock conditions):
\begin{eqnarray}
T_{\rm post\,shock} &=& \frac{3}{16}\,\frac{\mu\,m_P}{k_B}\,v_{\rm shock}^2 
\nonumber \\
&=& 0.105\,\textrm{keV}\,\left(\frac{v_{\rm shock}}{\textrm{300 km s$^{-1}$}}
\right)^2 
\end{eqnarray}
with $m_p$ the proton mass, $k_B$ the Boltzmann constant and $\mu=0.6$ the 
mean particle weight. Therefore the soft component temperature corresponds to 
a shock velocity of about 300 km s$^{-1}$, consistent with the expansion 
velocity of $260\pm50$ km s$^{-1}$ of the optical extended structure measured by
\citet{BBE03}. 

Furthermore the temperature of the hot component suggests the presence of 
another shock with a velocity of about 900 km s$^{-1}$. The spatial resolution 
is by far not good enough to further connect the apparent flow in HST 
observations with our results. However, if the optical extended structure 
represents a jet, then the second shock could be located at the base of the 
jet. The harder of the two X-ray components could also arise from the boundary 
layer between the accretion disk and the white dwarf. \citet{PCM05} found that
the maximum temperature in a cooling flow representing an optically-thin 
boundary layer is $k\,T_{\rm max} = 3/5\,k\,T_{\rm virial}$. The temperature 
inferred from a single-temperature fit will be lower than $k\,T_{\rm max}$. The 
plasma temperature from single temperature fits to boundary layer emission in 
\citet{KMS09} from RT Cru, T CrB, CH Cyg and SS73 17 are lower than the maximum
temperatures determined from cooling flow fits in RT Cru \citep{LuS07}, T CrB
\citep{LSM08} and SS73 17 \citep{ELS10}. The higher absorbing column in front 
of the hard component is consistent with this component being closer to the 
central engine than the lower temperature plasma.

The fact that the soft component of the X-ray spectrum is consistent with no 
internal absorption can only be explained if the source of the X-ray emission 
lies outside of the symbiotic nebula with a typical size of a few AU. In 
contrast, almost all symbiotic stars detected in X-rays show spectra absorbed 
by columns densities of a few time $10^{21}$ to $10^{23}$ cm$^2$ 
\citep[e.g.][]{LuS07, KMS09}. 

The over-abundance of N by a factor of $\approx 4$ has some implications for 
the origin of the shocked material. \citet{ScS90} find in V1329 Cyg an 
enhancement factor of about 10 in nitrogen in the symbiotic nebulosity from 
optical, NIR and UV spectroscopy. The mean abundances found in M giants show an 
over-abundance of N by a factor of 3.5 compared to solar values 
\citep{SmL85,SmL86}, due to the conversion of carbon into nitrogen and mixing 
into the atmosphere of the red giant \citep{NSS88}. Larger values may be 
attributed to more advanced nuclear burning stages. Therefore the shocked 
material in the softer component was probably initially part of the red giant 
wind, which has not been further reprocessed during thermonuclear burning close 
to the white dwarf \citep{ScS90} and has been ejected outside of the symbiotic 
nebula within a jet.

The possible lack of short term variability in the X-ray light curve may 
support the above scenario. Variability with time scales of minutes to hours is 
typically observed in symbiotics where X-ray originate in the accretion flow 
\citep[e.g. SS73 17, Eze et al. 2010, or CH Cyg,][]{MIK07}. However, because the
X-ray count rate from V1329\,Cyg leads to large error bars, we cannot 
definitely exclude flickering. Variability is statistically detected only in 
the UVW1 band at 2910 \AA{} and the rms amplitude of this variability is only 
1\%. 

\section{Conclusion} \label{sec_concl}

We report the detection for the first time of X-ray emission from the 
symbiotic star V1329 Cyg which was not previously detected by {\em ROSAT} due 
to its low X-ray flux. The observed X-ray temperatures and especially
the unabsorbed nature of the soft component of the X-ray spectrum indicate that 
some of the X-ray emission might originate in shocks inside a jet outside the 
symbiotic nebula. In this regard, V1329 Cyg adds to the small but growing group 
of jet-driving symbiotics with X-ray emission.

\acknowledgements
MS wishes to thank Hans Martin Schmid for fruitful discussions. This work is 
based on observations obtained with {\em XMM-Newton}, an ESA science mission 
with instruments and contributions directly funded by ESA Member States and the 
USA (NASA). We thank NASA for funding this work through XMM-Newton AO-8 awards 
NNX09AP88G to GJML and JLS, and NNX10AK31G to JLS. We also acknowledge helpful 
comments and suggestions by an anonymous referee.

\end{document}